\documentstyle[]{mn}
\newif\ifAMStwofonts

\def\etal{{\rm et al.}}

\def\simgt{\mathrel{\spose{\lower 3pt\hbox{$\sim$}}
        \raise 2.0pt\hbox{$>$}}}
\def\simlt{\mathrel{\spose{\lower 3pt\hbox{$\sim$}}
        \raise 2.0pt\hbox{$<$}}}

\ifoldfss
  \ifCUPmtlplainloaded \else
    \NewTextAlphabet{textbfit} {cmbxti10} {}
    \NewTextAlphabet{textbfss} {cmssbx10} {}
    \NewMathAlphabet{mathbfit} {cmbxti10} {} 
    \NewMathAlphabet{mathbfss} {cmssbx10} {} 
  \fi
  \ifAMStwofonts
    \ifCUPmtlplainloaded \else
      \NewSymbolFont{upmath} {eurm10}
      \NewSymbolFont{AMSa} {msam10}
      \NewMathSymbol{\upi}     {0}{upmath}{19}
      \NewMathSymbol{\umu}     {0}{upmath}{16}
      \NewMathSymbol{\upartial}{0}{upmath}{40}
      \NewMathSymbol{\leqslant}{3}{AMSa}{36}
      \NewMathSymbol{\geqslant}{3}{AMSa}{3E}

      \let\leq=\leqslant \let\le=\leqslant
      \let\geq=\geqslant \let\ge=\geqslant
    \fi
  \fi
\fi 

\ifnfssone
  \newmathalphabet{\mathit}
  \addtoversion{normal}{\mathit}{cmr}{m}{it}
  \addtoversion{bold}{\mathit}{cmr}{bx}{it}
  \newmathalphabet{\mathbfit} 
  \addtoversion{normal}{\mathbfit}{cmr}{bx}{it}
  \addtoversion{bold}{\mathbfit}{cmr}{bx}{it}
  \newmathalphabet{\mathbfss} 
  \addtoversion{normal}{\mathbfss}{cmss}{bx}{n}
  \addtoversion{bold}{\mathbfss}{cmss}{bx}{n}
  \ifAMStwofonts
    \ifCUPmtlplainloaded \else
      \UseAMStwoboldmath
      \makeatletter
      \new@mathgroup\upmath@group
      \define@mathgroup\mv@normal\upmath@group{eur}{m}{n}
      \define@mathgroup\mv@bold\upmath@group{eur}{b}{n}
      \edef\UPM{\hexnumber\upmath@group}
      \new@mathgroup\amsa@group
      \define@mathgroup\mv@normal\amsa@group{msa}{m}{n}
      \define@mathgroup\mv@bold\amsa@group{msa}{m}{n}
      \edef\AMSa{\hexnumber\amsa@group}
      \makeatother
      \mathchardef\upi="0\UPM19
      \mathchardef\umu="0\UPM16
      \mathchardef\upartial="0\UPM40
      \mathchardef\leqslant="3\AMSa36
      \mathchardef\geqslant="3\AMSa3E

      \let\leq=\leqslant \let\le=\leqslant
      \let\geq=\geqslant \let\ge=\geqslant
    \fi
  \fi
\fi 

\ifnfsstwo
  \DeclareMathAlphabet{\mathbfit}{OT1}{cmr}{bx}{it}
  \SetMathAlphabet\mathbfit{bold}{OT1}{cmr}{bx}{it}
  \DeclareMathAlphabet{\mathbfss}{OT1}{cmss}{bx}{n}
  \SetMathAlphabet\mathbfss{bold}{OT1}{cmss}{bx}{n}
  \ifAMStwofonts
    \ifCUPmtlplainloaded \else
      \DeclareSymbolFont{UPM}{U}{eur}{m}{n}
      \SetSymbolFont{UPM}{bold}{U}{eur}{b}{n}
      \DeclareSymbolFont{AMSa}{U}{msa}{m}{n}
      \DeclareMathSymbol{\upi}{0}{UPM}{"19}
      \DeclareMathSymbol{\umu}{0}{UPM}{"16}
      \DeclareMathSymbol{\upartial}{0}{UPM}{"40}
      \DeclareMathSymbol{\leqslant}{3}{AMSa}{"36}
      \DeclareMathSymbol{\geqslant}{3}{AMSa}{"3E}

      \let\leq=\leqslant \let\le=\leqslant
      \let\geq=\geqslant \let\ge=\geqslant
    \fi
  \fi
\fi 

\ifCUPmtlplainloaded \else
  \ifAMStwofonts \else 
    \def\upi{\pi}
    \def\umu{\mu}
    \def\upartial{\partial}
  \fi
\fi

\title[A small source in Q2237+0305 ?]
  {A small source in Q2237+0305 ?}
\author[J. S. B. Wyithe et al.]
  {J.~S.~B.~Wyithe$^{1,2},$ 
  R.~L.~Webster$^1$, 
  E.~L.~Turner$^2$ \\
  $^1$ School of Physics, The University of Melbourne, Parkville, Vic, 3052, 
Australia\\
  $^2$ Princeton University Observatory, Peyton Hall, Princeton, NJ 08544, USA\\ 
 Email: swyithe@astro.Princeton.edu, rwebster@physics.unimelb.edu.au, elt@astro.Princeton.edu }
\date{Accepted. Received}
\pagerange{\pageref{firstpage}--\pageref{lastpage}}
\pubyear{1999}

\def\LaTeX{L\kern-.36em\raise.3ex\hbox{a}\kern-.15em
    T\kern-.1667em\lower.7ex\hbox{E}\kern-.125emX}

\begin{document}

\label{firstpage}

\maketitle

\begin{abstract}

Microlensing in Q2237+0305 between 1985 and 1995 (e.g. Irwin et al. 1989;
Corrigan et al 1991; $\O$stensen et al. 1996) has been interpreted in two
different ways. Firstly, the observed variations can be explained through
microlensing by stellar mass objects of a continuum source with dimensions
significantly smaller than the microlens Einstein Radius ($\eta_{o}$), but
consistent with that expected for thermal accretion discs (e.g. Wambsganss,
Paczynski \& Schneider 1990; Rauch \& Blandford 1991). However, other studies
have shown that models having sources as large as 5 $\eta_{o}$ can reproduce
the observed variation (Refsdal \& Stabell 1993; Haugan 1996). In this paper
we present evidence in favour of a small source. Our approach uses the
distribution of microlensed light-curve derivatives to place statistical
limits (as a function of source size) on the number of microlens Einstein
radii crossed by the source during the monitoring period. In contrast to previous analyses, our results are
therefore not dependent on an assumed time-scale. Limits on the source size
are obtained from two separate light-curve features. Firstly, recently
published monitoring data (Wozniak et al. 2000a,b; OGLE web page) shows large
variations ($\sim$.8-1.5 magnitudes) between image brightnesses over a period
of $\sim 700$ days or $\sim$15\% of the monitoring period. Secondly, the 1988
peak in the image A light-curve had a duration that is a small fraction
($\la$0.02) of the monitoring period. Such rapid microlensing rises and
short microlensing peaks only occur for small sources. We find that the
observed large-rapid variation limits the source size to be $<$0.2$\eta_{o}$
(95\% confidence). The width of the light-curve peak provides a stronger
constraint of $<$0.025$\eta_{o}$ (99\% confidence). The Einstein radius
(projected into the source plane) of the average microlens mass
$\langle m\rangle$ in Q2237+0305 is
$\eta_o\sim 10^{17}\sqrt{\langle m\rangle}\,cm$.
The interpretation that stars are responsible for microlensing in Q2237+0305
therefore results in limits on the continuum source size that are consistent
with current accretion disc theory. 

\end{abstract}


\section{Introduction}

The object Q2237+0305 (Huchra et al. 1985) comprises a source quasar at a redshift of $z=1.695$ that is gravitationally lensed by a foreground galaxy with $z=0.0394$ producing 4 resolvable images with separations of $\sim 1''$.  Each of the 4 images are observed through the galactic bulge, which has a microlensing optical depth in stars that is of order unity (e.g. Kent \& Falco 1988; Schneider et al. 1988; Schmidt, Webster \& Lewis 1998). In addition, the proximity of the lensing galaxy means that the effective transverse velocity may be high, yielding an expected microlensing event time-scale significantly shorter than that of other lensed quasars. The combination of these considerations make Q2237+0305 the ideal object from which to study microlensing. Indeed, Q2237+0305 is the only object in which cosmological microlensing has been directly confirmed (Irwin et.al 1989; Corrigan et.al 1991; Wozniak et al. 2000a,b).

Initially, this confirmation came in the form of a $\sim$0.2 magnitude brightening of image A with a rise-time of $\sim26$ days (Corrigan et al. 1991). Wambsganss, Paczynski \& Schneider (1990) found that, assuming a galactic transverse velocity of $\sim 600km\,sec^{-1}$, this rise could be explained by microlensing due to stellar masses of a source having dimensions much ($<0.01\,ER$) smaller than the microlens Einstein radius (ER) and therefore the typical caustic spacing. 

Spectral observations of subsequent microlensing also support the case for a source that is small with respect to the microlens ER. Lewis et al. (1998) determined the ratios of emission line equivalent widths relative to one image. They show that the ratios vary between images at a single epoch, and that the ratio for a single image (image A) varies between two different epochs. This spectral change is interpreted as being due to the different spatial extents of the continuum and emission line regions being differentially amplified due to microlensing, and suggests that the continuum region is smaller than the typical caustic separation. 

Refsdal \& Stabell (1993) proposed a model with a very low average microlens mass ($\sim 10^{-5}M_{\odot}$), and a source size consistent in physical size with the models of Wambsganss, Paczynski \& Schneider (1990). In this model the source size is several ER. Surprisingly, brightness variations as large as $\sim0.5$ magnitudes are predicted by this model. On the other hand Witt \& Mao (1994) note that smeared out light curves are produced which have trouble producing  asymmetric events such as that observed for image A (1989-90), although this interpretation was disputed by Haugan (1996). Very large sources, ($>7$ER) have been ruled out by Refsdal and Stabell (1997) from the $CIV$ line data of Lewis et al (1998). 

There are two important reasons to distinguish between the cases of a source that is large/small with respect to the microlens ER.
Firstly, much observational effort (OGLE collaboration; Lens Monitoring Project, Apache Point Observatory) is currently being directed towards monitoring of Q2237+0305 in the hope of observing a caustic crossing (the event time-scale is of order months rather than years due to the large distance ratio). As has been discussed by several authors (e.g. Grieger et al. 1988; Grieger et al. 1991; Agol \& Krolic 1999; Mineshige \& Yonehara 1999), the light curve of a straight-single caustic event contains information on the source geometry on nano-arcsecond scales, providing otherwise unobtainable resolution for observation of a quasars continuum region. The mode of analysis discussed by these authors is only valid in the case of a source that is much smaller than the typical caustic spacing. In addition, the differential magnification probed by these methods is much more significant for small sources.
Secondly, a good understanding of the microlensing parameters (e.g. galactic transverse velocity and mean microlens mass) are required for the successful analysis of a caustic crossing event. One means of obtaining such understanding is to interpret the monitoring data in terms of the microlensing rate (e.g. Lewis \& Irwin 1996; Wyithe, Webster \& Turner 1999, 2000b, (hereafter WWT99, WWT00b)). The rate is approximately independent of source size for small sources, a feature that can be used to remove one degree of freedom from the problem. 

This paper presents arguments that support the hypothesis of a source that has dimensions smaller than the ER. Sections \ref{model} and \ref{data_sec} describe the microlensing models and the collection of published monitoring data, while section \ref{vels} describes how the distribution of microlensed light-curve derivatives can be used to place limits on the length of caustic network sampled by the observations. In section \ref{size} we discuss how two different light-curve features, in combination with the sampling length, limit the continuum source to be significantly smaller than the microlens ER. This paper differs from previous analyses by concentrating on determining the ratio of the source size to the microlens mass, rather than the source size in physical units. A more important difference is that, unlike previous work, our analysis contains no assumptions about the mean microlens mass or the transverse velocity (other than a prior probability for transverse velocity which, within reasonable limits, has no effect on our results (WWT00b)).

\section{The Microlensing model}
\label{model}

\begin{table}
\begin{center}
\caption{\label{params}Values of the total optical depth ($\kappa$) and the magnitude of the shear ($\gamma$) at the position of each of the 4 images of Q2237+0305. The quoted values are from Schmidt, Webster \& Lewis (1998).}
\begin{tabular}{|c|c|c|}
\hline
Image & $\kappa$   & $|\gamma|$  \\ \hline
  A   & 0.36                             &  0.40        \\
  B   & 0.36                             &  0.40        \\
  C   & 0.69                             &  0.71        \\
  D   & 0.59                             &  0.61        \\ \hline
\end{tabular}
\end{center}
\end{table}

 To model microlensing in Q2237+0305 we assume the macro-parameters for the lensing galaxy calculated by Schmidt, Webster \& Lewis (1998). These values are shown in table \ref{params}. 
 Two orientations were chosen for the transverse velocity with respect to the galaxy, with the source trajectory being parallel to the A$-$B and C$-$D axes.
 Photometric errors were simulated by applying a random perturbation to the model light-curves having a magnitude distributed according to a Gaussian of halfwidth $\sigma$. The simulations used two different estimates of the error in the photometric magnitudes. In the first case a small error was assumed (SE). For images A and B, $\sigma_{SE}$=0.01 mag, and for images C and D $\sigma_{SE}$=0.02 mag. In the second case, a larger error was assumed (LE). For images A and B, $\sigma_{LE}$=0.02 mag and for images C and D $\sigma_{LE}$=0.04 mag. The observational error in Irwin et al. (1989) was 0.02 mag. These models produce quantitatively similar results. Therefore, in this paper we present only results from models with small errors (SE) and a transverse velocity whose direction lies along the A-B axis.

 Both the microlensing rate due to a transverse velocity (e.g. Witt, Kaiser \& Refsdal 1993; Lewis \& Irwin 1996), as well as the corresponding rate due to random proper motions (Wyithe, Webster \& Turner 2000a (hereafter WWT00a)) are not functions of the details of the microlens mass distribution, but rather are dependent only on the mean microlens mass. The independence of the microlensing rate on the form of the mass-function has been checked for models containing mass ranges over 2 orders of magnitude. We assume that the dominant contribution to the optical depth comes from objects with masses differing by less than 2 orders of magnitude and consequently limit our attention to models in which all the point masses have the same mass since the results obtained will be applicable to other models with different forms for the mass function. In addition, we consider only models that contain no continuously distributed matter.

The distribution of microlensed light-curve derivatives has nearly the same form (up to a scaling factor in the derivative) whether the variation results from a transverse velocity of a static screen of stars, or whether it is due to their random proper motion (WWT00a). This suggests that microlensing due to the combination of a transverse velocity and stellar proper motions can be approximated using a static screen of microlenses with an effective transverse velocity (which is larger than the physical galactic transverse velocity), thus the microlensing effect of microlens proper motions is included implicitly (WWT99). In this paper we use model light-curves obtained using this effective transverse velocity approximation.

The ER of a microlens in the source plane is denoted by $\eta_{o} $.
 We consider Gaussian sources with sizes $S$ (where $S$ is the half width of the Gaussian intensity profile) ranging over three orders of magnitude, from $S=0.0015\eta_{o}-1.6\eta_{o}$. 
To construct microlensing light-curves for $S\leq0.05\eta_{o}$ we use the inversion technique of Lewis et al. (1993) and Witt (1993). For $S\geq0.1\eta_{o}$ we use the ray-tracing method (e.g. Kayser, Refsdal \& Stabell 1986; Schneider \& Weiss 1987; Wambsganss, Paczynski \& Katz 1989).

\subsection{Models for small sources}
The microlensing models for sources of size $S\leq0.05\eta_{o}$ presented in this work have been discussed in detail in WWT00b. Finite source light-curves were produced by convolving a Gaussian source profile with a point-source light-curve (e.g. Witt \& Mao 1994). For each combination of microlensing parameters, 100 light-curves of length 10$\eta_{o}$ were produced. 

\subsection{Models for large sources}
For source sizes $S\geq0.1\eta_{o}$, the 1-D approximation to the finite source light-curve is no longer valid, and the lower resolution requirements make ray-tracing the appropriate method. For sources having $0.1\eta_{o}\leq S\leq0.4\eta_{o}$, light-curves were produced from magnification maps having side-lengths of $40\eta_{o}$. For sources with sizes $0.8\eta_{o}\leq S\leq1.6\eta_{o}$, we produced magnification maps having side-lengths of $160\eta_{o}$. The magnification maps had a resolution of 500$\times$500 pixels, and the number of stars used in the models was calculated through the method described in Lewis \& Irwin (1995) and Wyithe \& Webster (1999). Finite source light-curves are produced from these maps through convolution with a 2-d Gaussian source profile.

\section{Monitoring data}
\label{data_sec}
\begin{figure*}
\vspace*{175mm}
\includegraphics{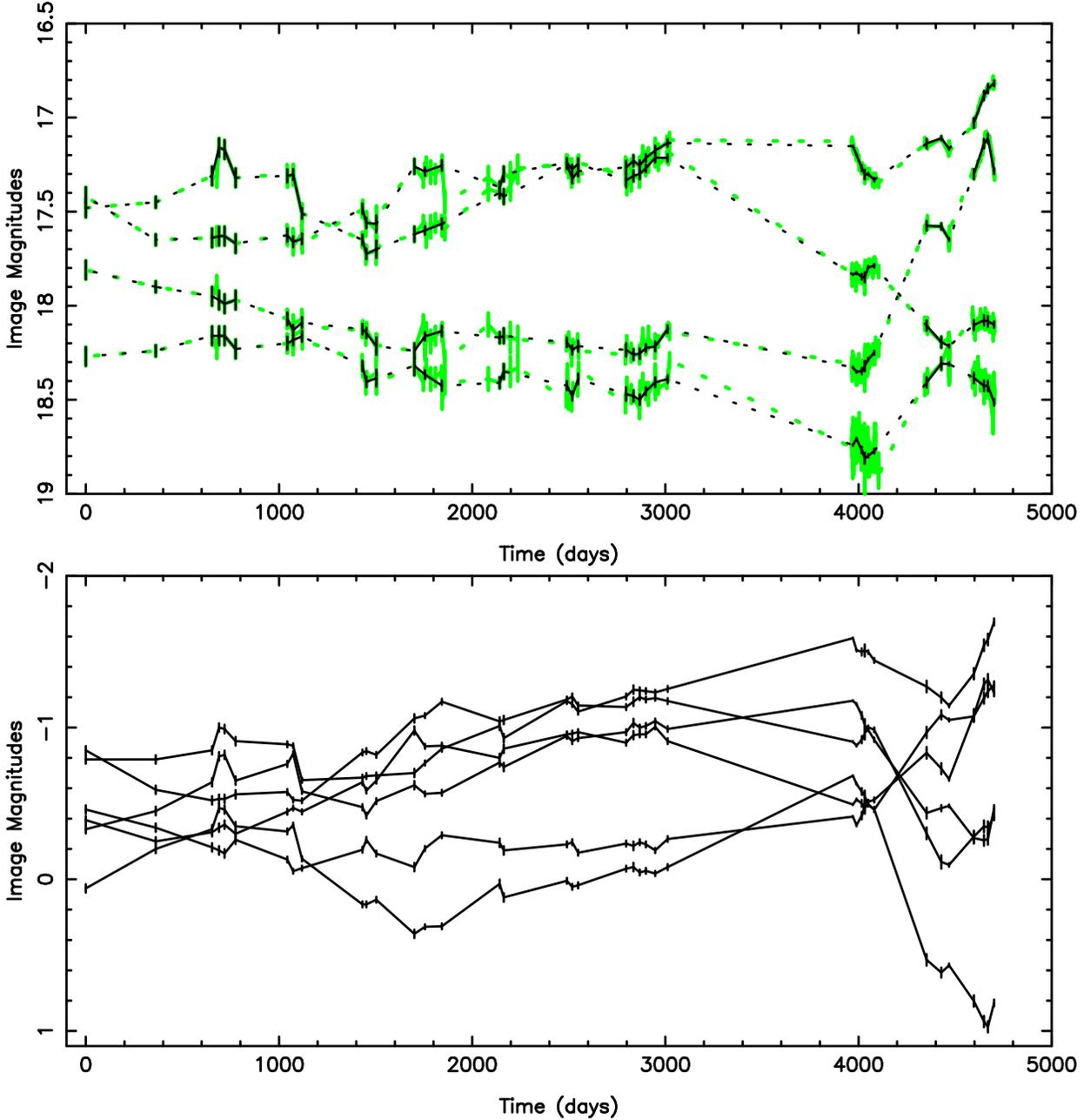}
\caption{\label{data}Compilation of published light-curve data for Q2237+0305. Top: Individual image light-curves. The light lines show the entire data set, and the dark lines show the modified data-set used for our analysis. Dotted lines span large gaps in the observations. Bottom: Six difference light-curves (A-B, A-C, A-D, B-C, B-D, C-D) calculated from the modified data set.}
\end{figure*}

We have compiled a data set that includes the photometry presented in Schneider et al. (1988), Kent \& Falco (1989), Irwin et al. (1989), Corrigan et al. (1991), $\O$stensen et al. (1996); Wozniak et al. (2000a,b) and the OGLE web page (see http://www.astro.princeton.edu/$\sim$ogle/ogle2/huchra.html). To successfully analyse the data, even sampling and accurate photometry are required so that light-curve derivatives at one or a few epochs do not dominate the statistics. We have therefore averaged points on the light-curves, producing a more even sampling rate and reducing photometric error according to the procedure described in WWT99. Points were averaged if the observations were taken within one week, a procedure that does not smooth out observed short fluctuations (less than 1 week) since these are not present in the data above the noise generated by photometric uncertainty. Following this, any points having an associated error above $\Delta M=0.05$ magnitudes were removed from the sample since data points with large errors substantially degrade the measurement of microlensing rate through introduction of noise into the low derivative regime. There were also two data points that were discarded since they displayed correlated flux variation in two images. The upper plot in figure \ref{data} shows the entire data set as well as the collection of points used for our analysis, dotted sections highlight large gaps between observations. Error bars are shown representing the published errors. Errors for averaged points have been added in quadrature. 

While data taken prior to $\sim 3100$ days is in $R$ and $r$-bands, observations from OGLE have been made in $V$-band. We have not included a colour correction since our analysis considers the difference between image magnitudes. In addition, the continuum at different wavelengths may be emitted from regions having different physical scales. Differential magnification due to microlensing therefore introduces a complication when comparing data taken in different bands. However, microlens induced colour change occurs primarily during caustic crossings (Wambsganss \& Paczynski 1991). Therefore if $S\ll\eta_o$, colour change is only observed for a small fraction of time, while if $S\sim\eta_o$ the differences are small. The lower plot of figure \ref{data} shows the set of difference light-curves on which our analysis is performed. We have drawn solid lines between all points on this plot since the derivatives between all points are used in the analysis.

\section{The sampling length}
\label{vels}

\begin{figure*}
\vspace*{210mm}
\includegraphics{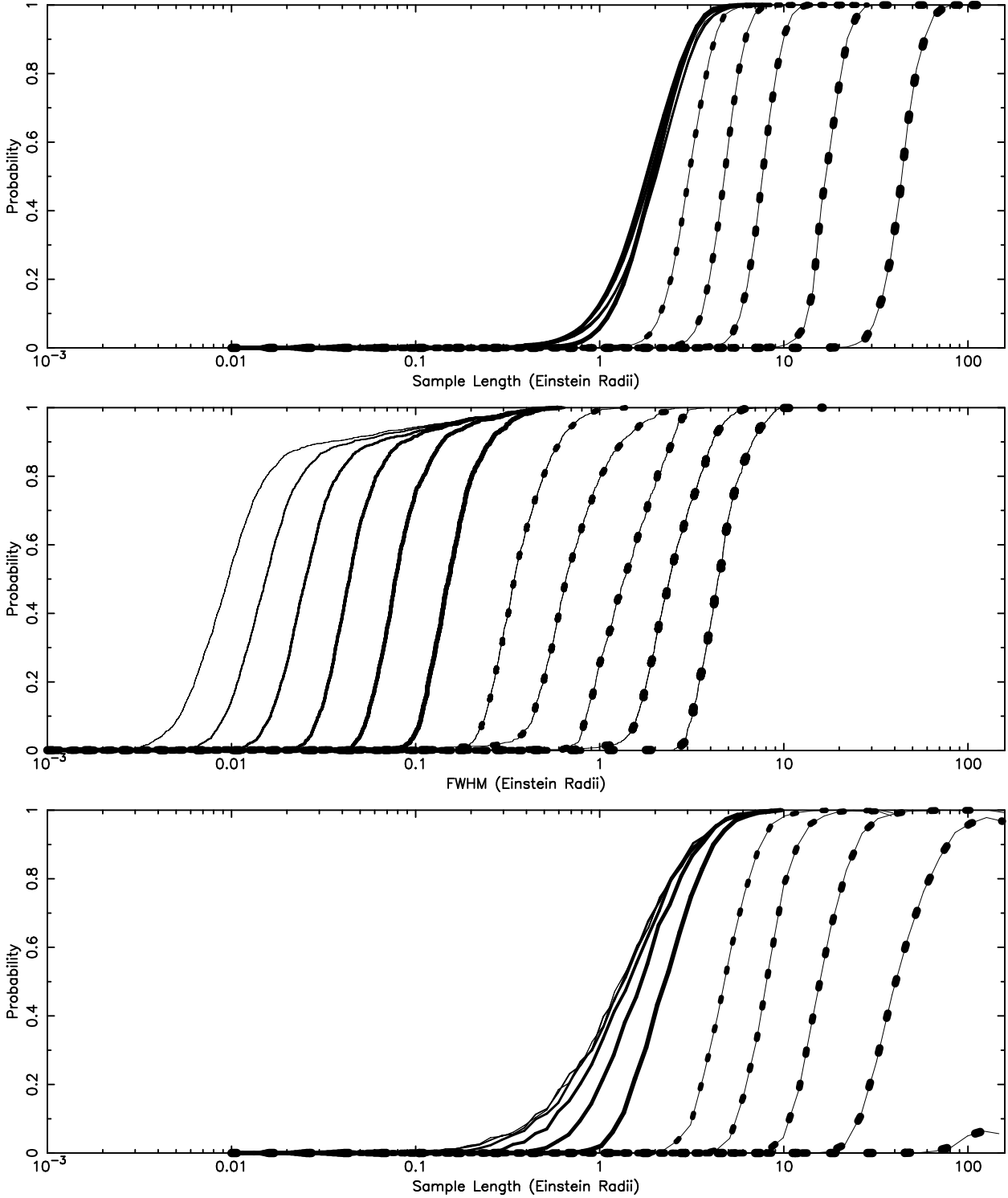}
\caption{\label{sample_length}Top: The probability $P_{\eta}(\eta<\eta_{period}|S)$ that the source has crossed less than $\eta_{period}$ ER during the monitoring period. Centre: The cumulative probability $P_{W}(W<W_o|S)$ for light-curve peak width. Bottom: The cumulative probability $P_m(\{\Delta M\})>\{\Delta M_{obs}\}|S,\eta_{period})$ for variation in a mock data set being larger than the observed variation. The probabilities are shown for $S=$ 0.0015625 0.003125, 0.00625 0.0125, 0.025, 0.05, 0.1, 0.2, 0.4, 0.8, 1.6 $\eta_o$ (Thicker lines denote larger sources). The solid and dotted lines correspond to sources with $S\leq0.05\eta_o$ and $S\geq0.1\eta_o$ respectively.}
\end{figure*}

WWT99,WWT00b describe a method to compute the probability for the value of the lensing galaxies effective transverse velocity using the distribution of microlensed light-curve derivatives. The effective transverse velocity obtained is in units of $km\,sec^{-1}\,\langle m\rangle^{-\frac{1}{2}}$ where $\langle m\rangle$ is the mean microlens mass. When multiplied by the length of the monitoring period, this yields the length of caustic structure sampled ($\eta_{period}$) in units of the ER of the average microlens mass ($\eta_{o}$). This is determined as a function of source size $S$. Note that $\eta_{period}(S)$ is free of assumptions about time-scale and microlens mass. 
We have computed the probability $p_{\eta}(\eta_{period}|S)$ for source sizes $S=$ 0.0015625 0.003125, 0.00625 0.0125, 0.025, 0.05, 0.1, 0.2, 0.4, 0.8 and 1.6$\eta_o$. For $S<0.1\eta_o$ and $S\ge0.1\eta_o$, $p_{\eta}(\eta_{period}|S)$ was computed from 5000 and 1000 mock sets of monitoring data respectively at each of 50 assumed sampling lengths $\eta$. The cumulative distributions 
\begin{equation}
P_{\eta}(\eta<\eta_{period})=\int_0^{\eta_{period}}\,p_{\eta}(\eta'|S)\,d\eta'
\end{equation}
are shown in the upper plot of figure \ref{sample_length} (thicker lines indicate larger sources). For $S\leq0.05\eta_{o}$ (solid lines), the measurement of sample length is approximately independent of source size (this independence is discussed in WWT99). However, the measured sample length increases as larger source sizes are assumed because larger sources have decreased event amplitudes and increased event time-scales. This combination results in a higher transverse speed being required to reproduce the observed derivatives. The limits on sample length are $\sim 5\eta_{o}$ for small sources, and $\sim 100\eta_{o}$ for all sizes considered.

\section{Limits on source-size}
\label{size}

In this section we place limits on the source size in units of microlens ER by combining sampling length limits with features in the monitoring light-curves. There are at least two features in the monitoring data that suggest a small source size. Firstly, while the 0.2 magnitude variation in image A (1989) can be explained by a large source, the associated peak has a very short duration with respect to the monitoring period ($\sim2\%$). The sample length ($\eta_{period}(S)$) must therefore be greater than $50\,S$ regardless of whether the peak is due to the source passing outside of a cusp or to a caustic crossing (the minimum event width is set by the source size in either case). We show below that a sample length of $50\,S$ can only be reconciled with a small source. Secondly, new monitoring data (Wozniak et al. 2000a,b; OGLE web page (see http://www.astro.princeton.edu/$\sim$ogle/ogle2/huchra.html)) shows large scale brightness variation in all images, with values ranging from $\sim$.5-1.25 magnitudes over a period of $\sim$700 days. While we find that (surprisingly) such large microlensing variation can be produced by large sources over long time-scales, rapid large scale microlensing variation only occurs for small sources. The quantitative analysis of the limits imposed on the source size by these features are discussed in turn. 

We note that choosing light-curve features a posteriori, and comparing them statistically to a sample of models in order to draw conclusions regarding input parameters (such as the source size) has the potential to introduce a bias. However in the present case we feel that our method is justified because of the well established prior knowledge that smaller microlensed sources produce larger, more rapid variations.

\subsection{Source-size limits from the short peak}

\begin{figure*}
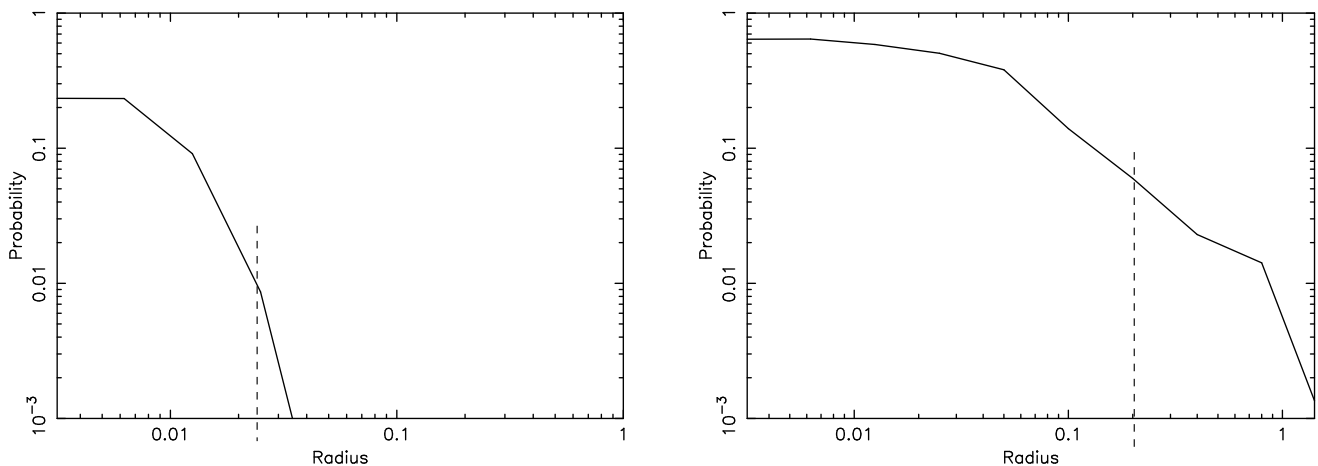

\vspace*{70mm}
\includegraphics{fig3a.epsi}
\includegraphics{fig3b.epsi}
\caption{\label{rad_prob}Left: The probability $P(R<R_{obs}|S)$ of observing a peak width to sample-length ratio smaller than that observed ($R_{obs}$) as a function of source size. The dashed line shows the 99\% confidence limit. Right: The probability $P(\{\Delta M\}>\{\Delta M_{obs}\}|S)$ of observing changes in the difference light-curves of $\Delta M^{i}>\Delta M_{obs}^{i}$ magnitudes for all $i$ during a period of $0.15\eta_{period}$ as a function of source-size. The dashed line shows the 95\% confidence limit.}
\end{figure*}

The 1988 peak in the light-curve of image A has been interpreted as one half of a double horned event (e.g. Racine 1992), though the data are not conclusive on this point. It had a height $\ga0.2$ magnitudes (the points measure the lower limit), and a duration of $\la100$ days (the available points measure an upper limit). The duration of the peak is $\sim 2.1\%$ of the total monitoring period ($\sim 4700$ days). We calculate the likely-hood of this duration for different source sizes. 

Firstly, we search for peaks in model light-curves (we do not demand that a peak be due to a caustic crossing) with maxima more than 0.2 magnitudes above the associated minima on both sides. Our peak statistics were obtained from sampling at the resolution of the simulation ($>$1 point per day). The sample of peaks for sources with $S\le0.05\eta_o$ was computed from 1000$\eta_o$ of light-curve. For $S\ge0.10\eta_o$, a sample of 1000 light-curve peaks was obtained. The central plot of figure \ref{sample_length} shows the cumulative distributions $P_W(W<W_o|S)$ of peak full-width-at-half-maximums ($W$) given source-size $S$ (sizes assumed are the same as those in section \ref{vels}). $W$ is always larger than $S$ and may be 10-100 times larger. For smaller sources the distribution of $W$ has two components. The typical $W$ for a peak resulting from a caustic crossing is approximately proportional to $S$. However, if the peak is the result of the source having passed outside a cusp then the ratio $W:S$ may be arbitrarily large, since in this case $W$ is approximately independent of $S$ for $S\la0.05\eta_o$. For larger sources, $W$'s for caustic crossing events are typically as large as those for cusp related events, while for $S\sim \eta_o$ no classification is possible since the source is generally in contact with two or more caustics.  

 From model light-curves we calculate the ratios of $W$ to $\eta_{period}$. The probability $P$ of finding a ratio   
\begin{equation}
R(S)=\frac{W(S)}{\eta_{period}(S)}
\end{equation}
smaller than that observed ($R_{obs}$) as a function of source size is
\begin{eqnarray}  
\nonumber
&&\hspace{-5mm}P(R<R_{obs}|S)=\\
&&\hspace{-7mm}\int_{0}^{W_{max}}\hspace{-1mm}\int_{\eta_{period}'=\frac{W}{R_{obs}}}^{\eta_{max}} p_w(W'|S)p_{\eta}(\eta_{period}'|S)d\eta_{period}'dW',
\end{eqnarray}
where $p_w=\partial P_W/\partial W$. $W_{max}$ and $\eta_{max}$ are the largest values for $W$ and $\eta_{period}$ considered in the simulations and have associated probabilities of 0.
The left-hand plot of figure \ref{rad_prob} shows the probability $P(R<R_{obs}|S)$. We find that the short duration of the light-curve peak relative to the monitoring period limits the source size $S$ to be smaller than $\sim 0.025\eta_o$ at the 99\% level.

\subsection{Source size limits from large, rapid microlensing variability}

In this section we consider observed large scale variation over a period of $\sim$2 years (1997-1999). On this time-scale, intrinsic source variation may be an important factor in the absolute brightness variation of a single image. We therefore consider the relative change between image magnitudes, which due to the short time-delay (of order 1 day (e.g. Schneider et al. 1988)) is approximately independent of intrinsic variation.  
The maximum magnitude changes in difference light-curves over periods $\leq700$ days ($\sim15\%$ of the sample length) are $\Delta M_{A-B}$=0.83, $\Delta M_{A-C}$=0.92, $\Delta M_{A-D}$=1.00, $\Delta M_{B-C}$=1.50, $\Delta M_{B-D}$=0.91 and $\Delta M_{C-D}$=0.96. We define a set of observed magnitude variations in ascending order: $\{\Delta M_{obs}\}=\{0.83, 0.91, 0.92, 0.96, 1.00, 1.50\}$. For each assumed monitoring period, 5000 (for $S\le0.05\eta_o$) or 1000 (for $S\ge0.10\eta_o$) mock data sets were calculated using a sampling rate with identical relative spacing to the monitoring data. For each simulated data set we calculate $\{\Delta M\}$ such that $\Delta M^{i}>\Delta M^{i-1}$ for $2\leq i\leq6$, thus we do not specify between which pairs of images the observed changes should be seen. We calculate the probability 
\begin{equation}
P_m(\{\Delta M\}>\{\Delta M_{obs}\}|S,\eta_{period})
\end{equation}
 of finding a change in the difference light-curves of $\Delta M^{i}>\Delta M_{obs}^{i}$ magnitudes for all $i$ over $0.15\eta_{period}$ (during any part of the period). The distributions are shown in the lower panel of figure \ref{sample_length}. Larger sources are less likely to exhibit the observed variation over $\sim$15\% of a given sample length. For example, while model source sizes of $S\la0.05\eta_{o}$ always exhibit the observed level of microlensing variation over sample lengths $\ga10\eta_o$, the $S=1.6\eta_o$ source attains these values less than 3\% of the time over a sample length of $\sim150\eta_o$.

$P_m(\{\Delta M\}>\{\Delta M_{obs}\}|S,\eta_{period})$ is convolved with $p_{\eta}(\eta_{period}|S)$ to find
\begin{eqnarray}
\nonumber
&&\hspace{-6mm}P(\{\Delta M\}>\{\Delta M_{obs}\}|S)=\\
&&\hspace{-7mm}\int\,P_m(\{\Delta M\}>\{\Delta M_{obs}\}|S,\eta_{period})\,p_{\eta}(\eta_{period}|S)\,d\eta_{period}.
\end{eqnarray}
This function is shown in the right-hand panel of figure \ref{rad_prob}. We find that the large, rapid changes in the image magnitudes limit the source to be smaller than $0.2\eta_o$ at the 95\% level. 

We note that the calculation of the probabilities $P(\{\Delta M\}>\{\Delta M_{obs}\}|S)$ assumes that $P_m(\{\Delta M\}>\{\Delta M_{obs}\}|S,\eta_{period})$ and $p_{\eta}(\eta_{period}|S)$ are independent. This is a false assumption since the two functions have been calculated from the same data set. However, the observation of the large scale variation (in the OGLE data) introduces large derivatives which increases the estimate of sample length (over the estimate that would be made from the pre 1996 data alone, scaled by the relative monitoring periods). In the case of a longer sample length, a source has more caustic network over which to undergo large-scale microlensing variation. This results in the inference of a larger upper limit on source size than would be expected if the sampling length were deduced from the pre-1996 data alone. Any bias introduced by the co-dependence of $P_m(\{\Delta M\}>\{\Delta M_{obs}\}|S,\eta_{period})$ and $p_{\eta}(\eta_{period}|S)$ therefore results in more conservative (ie. larger) source-size limits than the case where the sample length and large-scale variation were measured from different data-sets.

\section{Conclusion}

We have used the distribution of microlensed light-curve derivatives to find probability functions for the length of caustic structure sampled by monitoring observations of Q2237+0305 as a function of assumed source size. The 1988 light-curve peak had a height larger than 0.2 magnitudes, and a duration less than 0.02 of the monitoring period. At the 99\% level such a short peak can only be explained for a source with dimensions smaller than $\sim$0.025$\eta_{o}$. In addition, monitoring by OGLE shows changes in difference light-curves ranging from $\sim0.8-1.5$ magnitudes over $\sim$15\% of the monitoring period. We find that such rapid large scale changes can only be explained by a source that is smaller than $\sim0.2\eta_{o}$ (95\% confidence). Importantly, these limits are independent of any assumption about mean microlens mass or galactic transverse velocity.

The Einstein radius of the average microlens mass $\langle m\rangle$ in Q2237+0305 is $\eta_o\sim 10^{17}\sqrt{\langle m\rangle}\,cm$. In combination with our limits in $S$, the assumption of stellar mass microlenses therefore impose a limit of $\la 2\times10^{15}-2\times10^{16}\,cm$ on the continuum source size (consistent with the typical scale-size expected for a continuum emitting accretion disc about a super-massive black hole (e.g. Rees 1984; Rauch \& Blandford 1991; Agol \& Krolik 1999). Conversely, if a source size of $\sim 10^{15}cm$ is assumed, then the our limits on microlens mass corresponding to the short light-curve peak and large microlens variation are respectively $\langle m\rangle>0.25M_{\odot}$ (99\%) and $\langle m\rangle>0.0025M_{\odot}$ (95\%). Note that for a source having a diameter of $ \sim 10^{15}cm$ we have explicitly checked masses down to $\langle m\rangle\sim 10^{-5}M_{\odot}$. A much smaller source must therefore be assumed if sub-stellar masses were to form the bulk of the galactic bulge in Q2237+0305.

Our calculations assume that the dominant contribution to optical depth comes from a population of microlenses having a range smaller than two orders of magnitude. We note that our model therefore does not account for the possibility suggested by Refsdal \& Stabell (1993) in which a population of stars and another of very low-mass objects both contribute significantly to the optical depth. In this case, the rapid variation due to the small masses, and the slower variation due to the large masses should be approximately superimposed (Refsdal \& Stabell 1993). However our results can be qualitatively interpreted for this scenario. There are two possibilities. Firstly, if the source has dimensions $S\ll\eta_{sm}$ (where $\eta_{sm}$ is the Einstein radius of the mean microlens mass of the low mass population), then our results hold since $\langle m\rangle$ is quite insensitive to the value of the heavy masses. Secondly, if $S\gg\eta_{sm}$ then the light-curves will be like those resulting from microlensing by the large masses only (with a continuous component of mass density), combined with a continuous low amplitude flicker. In this case, the results we have presented can be interpreted as the ratio of source size to the Einstein radius of the mean microlens mass of the large population. We therefore conclude that a significant contribution to the optical depth comes from objects with Einstein radius larger than the source size, regardless of the mass function.

 The conclusion that the source is small with respect to the microlens Einstein radius is important because it means that caustic crossing light-curves can be inverted to obtain extremely high resolution information on the continuum source structure.

\section{acknowledgements}
The authors would like to thank Sjur Refsdal and Joachim Wambsganss for enlightening discussions, as well as the anonymous referee for suggestions which have enhanced the presentation of this work. We would also like to acknowledge the OGLE collaboration for making their monitoring data publically available in real time. This work was supported by NSF grant AST98-02802. JSBW acknowledges the support of an Australian Postgraduate Award and a Melbourne University Postgraduate Overseas Research Experience Award.

\label{lastpage}

\end{document}